\documentclass[12pt]{article}

\usepackage[T1]{fontenc}
\usepackage{lmodern}
\usepackage[final]{microtype}
\usepackage{amsmath,amssymb,graphicx}
\usepackage[hidelinks]{hyperref}
\usepackage{xurl}
\usepackage{abstract}
\usepackage[a4paper, margin=1in]{geometry}
\usepackage[x11names, dvipsnames]{xcolor}
\usepackage{booktabs}
\usepackage[font=small,labelfont=bf]{caption}
\usepackage{float}
\usepackage[section]{placeins}
\usepackage{enumitem}
\usepackage{titlesec}

\titlespacing*{\section}{0pt}{1.4ex plus 0.2ex minus 0.1ex}{0.6ex}
\titlespacing*{\subsection}{0pt}{0.9ex plus 0.15ex minus 0.08ex}{0.35ex}
\titlespacing*{\paragraph}{0pt}{0.35ex plus 0.05ex minus 0.05ex}{0.35em}

\titlespacing*{\section}{0pt}{0.8ex plus 0.15ex minus 0.1ex}{0.35ex}
\titlespacing*{\subsection}{0pt}{0.55ex plus 0.1ex minus 0.08ex}{0.22ex}
\titlespacing*{\paragraph}{0pt}{0.22ex plus 0.05ex minus 0.05ex}{0.28em}

\setlist[itemize]{topsep=1pt,itemsep=0pt,parsep=0pt,partopsep=0pt,leftmargin=1.2em}
\setlist[enumerate]{topsep=1pt,itemsep=0pt,parsep=0pt,partopsep=0pt,leftmargin=1.4em}

\begin{document}

\begin{center}
{\Large\bfseries Polynomial Multiproofs\\[0.08em]
for Scalable Data Availability Sampling\\[0.08em]
in Blockchain Light Clients\par}
\vspace{0.4em}
{\normalsize Rachit Anand Srivastava\textsuperscript{*}, Vikram Bhattacharjee\textsuperscript{*}, Will Arnold, Toufeeq Pasha\par}
\vspace{0.12em}
{\normalsize Avail\par}
\end{center}

\vspace{0.2em}

\begingroup
\renewcommand\thefootnote{}
\footnotetext{\textsuperscript{*}Both authors contributed equally to this work.}
\addtocounter{footnote}{-1}
\endgroup

\begin{abstract}
\noindent
Light clients are essential for scalable blockchain systems because they verify data availability without downloading full blocks. In data-availability-sampling-based systems, sampled cells are retrieved from a peer-to-peer network and verified against cryptographic commitments. A common deployment pattern associates each sampled cell with an independent Kate--Zaverucha--Goldberg (KZG) proof, creating substantial cumulative bandwidth, storage, and verification overhead. This paper studies polynomial multiproofs (PMP) as a mechanism for reducing these costs in blockchain light clients. We present a design in which multiple sampled cell evaluations are verified using a single aggregated proof over a shared evaluation micro-domain and describe the corresponding changes to proof generation, dissemination, retrieval, and verification in a peer-to-peer light-client stack. We instantiate and evaluate the design in Avail, a modular data availability layer for blockchains, as a case study. The results show lower proof bytes, lower verifier CPU and memory usage, and deployment-level infrastructure cost reductions of up to 45\% relative to a per-cell baseline, while also clarifying the trade-offs introduced by grouped retrieval.
\end{abstract}

\section{Introduction}

\setlength{\parindent}{0pt}

Blockchain systems face a basic scalability problem: strong verification usually requires every node to download and validate full blocks, which imposes substantial communication, storage, and computation costs~\cite{albassam2018fraud,albassam2019lazyledger}. These costs are especially restrictive for light clients running on mobile devices, browsers, and edge environments~\cite{krol2023ethereumdas}. For light clients to be practical at scale, data availability verification must remain both sound and efficient.

A common approach is \emph{Data Availability Sampling} (DAS), in which block data is erasure coded into cells and clients sample a subset of coordinates to obtain probabilistic assurance that the full data is available~\cite{hallandersen2023foundations,albassam2018fraud}. Many modern deployments authenticate sampled cells using Kate--Zaverucha--Goldberg (KZG) polynomial commitments~\cite{kate2010kzg,hallandersen2023foundations}. In a baseline design, however, each sampled cell carries an independent proof, so the cumulative cost of proof transmission, storage, and verification grows quickly with the number of samples.

These costs are amplified in peer-to-peer settings, where cells and proofs are retrieved through a distributed hash table (DHT)~\cite{maymounkov2002kademlia,krol2023ethereumdas}. We address this bottleneck by studying \emph{polynomial multiproofs} (PMP) as a mechanism for reducing proof overhead in blockchain light clients~\cite{boneh2020multipoint}. Instead of verifying each sampled cell with a separate KZG proof, a light client verifies multiple sampled cells with one aggregated proof, while the peer-to-peer layer stores grouped proof-carrying objects rather than per-cell objects. We instantiate this design in Avail as a case study.

A key requirement is that aggregation must not weaken DAS security. In our design, grouping changes only the transport unit exposed by the DHT. The light client still reasons about sampled coordinates individually. To make this precise, we bind each aggregated proof to a fixed micro-domain of evaluations and to canonical grouping metadata, so the verifier knows exactly which opened values belong to each proof.

The main contributions of this paper are:
\begin{itemize}
    \item identifying per-cell KZG proofs as a bottleneck for DAS light clients in peer-to-peer settings and designing a multiproof retrieval and verification model for grouped sampled cells;
    \item specifying the implementation path from full-node proof generation to fat-client DHT publication and light-client verification; and
    \item implementing and evaluating the design in Avail, showing lower verifier overhead and up to 45\% lower infrastructure cost.
\end{itemize}

\section{Related Work}
\label{sec:related}

Data availability has emerged as a central problem in scalable blockchain verification. Al-Bassam et al.~\cite{albassam2018fraud} introduced fraud and data availability proofs for strengthening light-client security, and LazyLedger later proposed an architecture centered on ordering and data availability rather than global execution~\cite{albassam2019lazyledger}. Hall-Andersen et al.~\cite{hallandersen2023foundations} formalized DAS and analyzed its security in relation to erasure coding and commitment schemes. At the networking layer, Kr\'ol et al.~\cite{krol2023ethereumdas} studied the peer-to-peer requirements of Ethereum-style DAS.

Polynomial commitments provide the cryptographic basis for many DAS constructions. Kate, Zaverucha, and Goldberg introduced constant-size commitments with succinct evaluation proofs~\cite{kate2010kzg}. Boneh et al.~\cite{boneh2020multipoint} showed how multiple polynomial evaluations can be aggregated into a single proof. Our work differs from this prior literature by studying how such multiproofs can be integrated into a peer-to-peer light-client workflow, where proof dissemination, retrieval granularity, and verifier cost all matter at the systems level.

A relevant engineering reference point is Ethereum's EIP-4844 and the broader proto-danksharding effort, which deploy KZG commitments and batch verification in a large-scale production setting. That work focuses on validator-side blob verification, whereas we focus on peer-to-peer dissemination and reuse of proof-carrying objects for DAS light clients. In particular, validator-side systems often batch independent openings using pairing-product checks, which is an important comparison point for our evaluation.

Recent work on DAS networking has explored stronger dissemination and robustness models than a conventional DHT. PANDAS~\cite{pigaglio2025pandas} redesigns dissemination around direct, lightweight exchanges to meet consensus-time constraints, while Robust Distributed Arrays pursue stronger end-to-end robustness guarantees. Related measurement work on large peer-to-peer content-routing systems also helps contextualize the assumptions behind DHT-based lookup performance. For example, recent studies of IPFS report meaningful centralization and heterogeneity in peer quality, which can materially affect routing reliability, latency, and lookup behavior in practice~\cite{balduf2023ipfs}. Our work is complementary: rather than redesigning the dissemination substrate, we optimize the cryptographic and storage paths within a DHT-oriented light-client architecture.

\section{Polynomial Multiproof Construction}
\label{sec:construction}

A polynomial commitment scheme (PCS) is defined as a tuple of algorithms or protocols
\[
\mathcal{S} = (\textsf{gen}, \textsf{com}, \textsf{open}).
\]
Here \textsf{gen} outputs a structured reference string (SRS), \textsf{com} commits to a polynomial, and \textsf{open} proves correct evaluation on designated sets.

\subsection{Generic Multiproof Form}

The construction used in our case study follows the first scheme of Boneh et al.~\cite{boneh2020multipoint}, which is designed to optimize opening efficiency. To avoid notation overload, we use $U_i \subseteq T$ for the opened subset associated with polynomial $f_i$, where $T=\{z_1,\ldots,z_t\}$ is a common ambient evaluation domain.

The prover computes
\begin{equation}
h(X) := \sum_{i \in [k]} \gamma^{i-1} \cdot \frac{f_i(X) - r_i(X)}{Z_{U_i}(X)},
\label{eq:h-open}
\end{equation}
where $r_i$ is the interpolation polynomial matching the opened values on $U_i$ and $\gamma$ is the verifier challenge. The verifier checks one aggregated pairing equation rather than one equation per opening.

\subsection{Shared-Point Specialization Used Here}

To obtain a design suitable for grouped DAS retrieval, we use the shared-point specialization
\begin{equation}
U_i = T_g \quad \forall i \in [k],
\label{eq:same-set}
\end{equation}
where $T_g$ is a fixed micro-domain of size $g$ inside one row- or column-commitment context. Each opened polynomial still satisfies the same degree bound $d$, but all openings appearing in one aggregated proof are over the same micro-domain $T_g$.

Under this specialization,
\begin{equation}
h(X)=\frac{\sum_i \gamma^{i-1}\left(f_i(X)-r_i(X)\right)}{Z_{T_g}(X)}.
\label{eq:h-poly}
\end{equation}
The prover therefore computes the weighted polynomial sum, divides by $Z_{T_g}$, and discards the remainder. This avoids interpolation during proof generation.

The verifier computes
\begin{equation}
Z_i := [Z_{T_g \setminus U_i}(x)]_2 = [1]_2 = g_2,
\label{eq:zi}
\end{equation}
because $U_i=T_g$ for every opened polynomial, and then reduces the check to a single aggregated pairing relation:
\begin{equation}
\begin{aligned}
F &:= \prod_i e\left(\gamma^{i-1} \cdot (c_i - [r_i(x)]_1), Z_i \right) \\
  &= e\left(\sum_i \gamma^{i-1} c_i - \left[ \sum_i \gamma^{i-1} r_i(x) \right]_1, g_2\right).
\end{aligned}
\label{eq:F-expanded}
\end{equation}

\vspace{0.12em}
\begin{table}[htbp]
\centering
\small
\begin{tabular}{@{}lcc@{}}
\toprule
\textbf{Operation} & \textbf{Open} & \textbf{Verify} \\
\midrule
Polynomial interpolation & 0 & 1 \\
$\mathbb{G}_1$ scalar multiplication & $d+1-|T_g|$ & $N + |T_g| + 1$ \\
$\mathbb{G}_2$ scalar multiplication & 0 & $d$ (cacheable) \\
Pairing & 0 & 2 \\
\bottomrule
\end{tabular}
\caption{Operation summary for the shared-point specialization.}
\label{tab:ops}
\end{table}
\vspace{0.12em}

Table~\ref{tab:ops} summarizes the opening and verification cost under the shared-point specialization. These counts apply to the specialized construction above, not to generic batch verification of independent single-point KZG openings.

\subsection{Mapping to DAS Grouping}

The main source of ambiguity in practice is the relationship between the transport group and the point set used by the verifier. In our instantiation, these are the same object: each grouped retrieval object corresponds to a full micro-domain
\[
T_g = \{z_{j_1},\ldots,z_{j_g}\}
\]
inside a single row or column context. The grouped object therefore transports all evaluations in that micro-domain for each included commitment context, rather than only an arbitrary subset of them. This is what makes the specialization in Equation~\ref{eq:same-set} correct.

Concretely, let a row polynomial $f_r$ be evaluated over a full row domain $T_{\mathrm{row}}$. The implementation partitions $T_{\mathrm{row}}$ into disjoint micro-domains of size $g$. For one such block
\[
T_g = \{z_{j_1},\ldots,z_{j_g}\} \subset T_{\mathrm{row}},
\]
the grouped object stores the entire value vector
\[
\left(f_r(z_{j_1}),\ldots,f_r(z_{j_g})\right)
\]
together with the commitment identifier and block metadata. If several rows are grouped together under one proof, each row contributes its full evaluation vector on the same micro-domain $T_g$.

\paragraph{Worked example.}
Suppose a row commitment context uses row domain
\[
T_{\mathrm{row}}=\{z_1,\ldots,z_8\},
\]
and the implementation chooses micro-domain size $g=4$. One grouped object may correspond to
\[
T_g=\{z_1,z_2,z_3,z_4\}.
\]
If two adjacent rows are packed together, then the object transports
\[
(f_r(z_1),f_r(z_2),f_r(z_3),f_r(z_4))
\]
and
\[
(f_{r+1}(z_1),f_{r+1}(z_2),f_{r+1}(z_3),f_{r+1}(z_4)).
\]
A light client may have sampled only one or two coordinates inside this group, but verification is performed against the full transported micro-domain $T_g$. This removes the earlier ambiguity between the sampled subset and the verifier's opened set.

\subsection{Non-Interactive Challenge Derivation}

Because aggregated proofs are stored in the DHT and reused by multiple clients, we instantiate the verifier challenge $\gamma$ using Fiat--Shamir:
\begin{equation}
\gamma = H(\mathsf{domain} \parallel \mathsf{srs\_id} \parallel \mathsf{cm}_1 \parallel \cdots \parallel \mathsf{cm}_k \parallel T_g \parallel \mathsf{coords} \parallel \mathsf{GCellBlock}).
\label{eq:fs-gamma}
\end{equation}
The transcript binds the challenge to the commitment context, the exact micro-domain, the grouping metadata, and the SRS domain. This prevents replay across different grouping layouts or heterogeneous commitment contexts, assuming collision resistance of the hash function and a common SRS for all commitments appearing in the proof.

\section{Instantiation in Avail}
\label{sec:integration}

We instantiate the design in Avail. Row and column commitments, together with the corresponding proof material, are generated by Avail full nodes. The implementation path is as follows.

\begin{enumerate}
    \item A full node computes the encoded block, row and column commitments, and the evaluation values required for each configured micro-domain.
    \item For each micro-domain, the full node constructs the aggregated proof input: commitment identifiers, full evaluation vectors over $T_g$, and block-position metadata.
    \item A fat client fetches authenticated headers and proof-carrying objects from one or more full nodes, packages them into grouped retrieval objects, and republishes them into the DHT.
    \item A light client samples coordinates, maps each sampled coordinate to its containing micro-domain, fetches the corresponding grouped object from the DHT, authenticates the commitment root from the header, and verifies the grouped proof.
\end{enumerate}

In the baseline design, each sampled data entry is stored together with an independent proof in the DHT, which duplicates proof metadata at fine granularity and increases the number of DHT objects. We therefore redesign the DHT layout so that a single proof authenticates multiple scalars grouped into a larger retrieval unit.

\subsection{Grouped Object Types}

The grouped retrieval path in Avail is represented by two core structures: \texttt{GCellBlock}, which identifies the covered region of the encoded block, and \texttt{MCell}, which packages the transported scalar values together with one aggregated proof.

\begin{figure}[htbp]
\centering
\fbox{%
\parbox{0.78\linewidth}{%
\ttfamily\small
struct GCellBlock \{\\
\hspace*{1.5em}rows\_start: u32,\\
\hspace*{1.5em}rows\_end: u32,\\
\hspace*{1.5em}cols\_start: u32,\\
\hspace*{1.5em}cols\_end: u32,\\
\}\\[0.4em]
struct MCell \{\\
\hspace*{1.5em}proof: [u8; 48],\\
\hspace*{1.5em}block: GCellBlock,\\
\hspace*{1.5em}count: u32,\\
\hspace*{1.5em}scalars: Vec<[u64; 4]>,\\
\}
}%
}
\caption{Core grouped retrieval structures used in the multiproof path.}
\label{fig:mcell-gcell-structs}
\end{figure}

Figure~\ref{fig:mcell-gcell-structs} shows the logical structure of the grouped retrieval object used by the multiproof path. \texttt{GCellBlock} binds the proof to a canonical block region, while \texttt{MCell} carries the proof, the covered region, and the transported scalar values.

\subsection{Threat Model and Grouping Trade-Offs}

Grouping preserves the DAS acceptance rule but brings additional network-level considerations. Grouped requests may expose interest in nearby coordinates, reducing unlinkability compared with per-coordinate retrieval. Likewise, withholding at the level of grouped objects can increase correlation among retrieval failures. These are trade-offs of the grouped retrieval design within the DHT-based model considered here.

The Fiat--Shamir transcript is bound to the SRS identifier, ordered commitment list, micro-domain $T_g$, coordinates, and \texttt{GCellBlock} metadata. Under canonical serialization, this binding helps prevent a proof constructed for one grouped object from being reused under a different grouping, ordering, or commitment context without changing the derived challenge.
\paragraph{Conservative soundness accounting.}
If a deployment originally budgets $s$ independent sampled coordinates under a per-cell retrieval model, then under worst-case perfect correlation inside groups of size $g$, a conservative lower bound on the number of effectively independent events is approximately $\lfloor s/g \rfloor$. To preserve the same target soundness level under this pessimistic model, one should therefore budget roughly $g$ times as many coordinate samples or, equivalently, ensure that at least the original target number of \emph{distinct groups} is sampled. This is intentionally conservative, but it makes the grouping-versus-soundness trade-off explicit.

\paragraph{Privacy trade-off.}
The same grouping factor $g$ also controls privacy leakage. Larger groups improve proof amortization and locality but reveal a larger neighborhood of queried coordinates to a serving peer. Smaller groups preserve more of the unlinkability of per-cell retrieval. In practice, $g$ should be chosen jointly with the sampling budget and the expected threat model.

\subsection{Encoding and Retrieval Model}

We use the following visual encoding:
\begin{itemize}
  \item Grid entry (32-byte data): {\color{Red} $\bullet$}
  \item Commitment: {\color{ForestGreen} $\bullet$}
  \item Cell retrieved through the DHT: {\color{black} $\Box$}
  \item Proof for that cell: {\color{Violet} $\bullet$}
\end{itemize}

\vspace{0.12em}
\begin{figure}[htbp]
    \centering
    \includegraphics[width=0.62\linewidth]{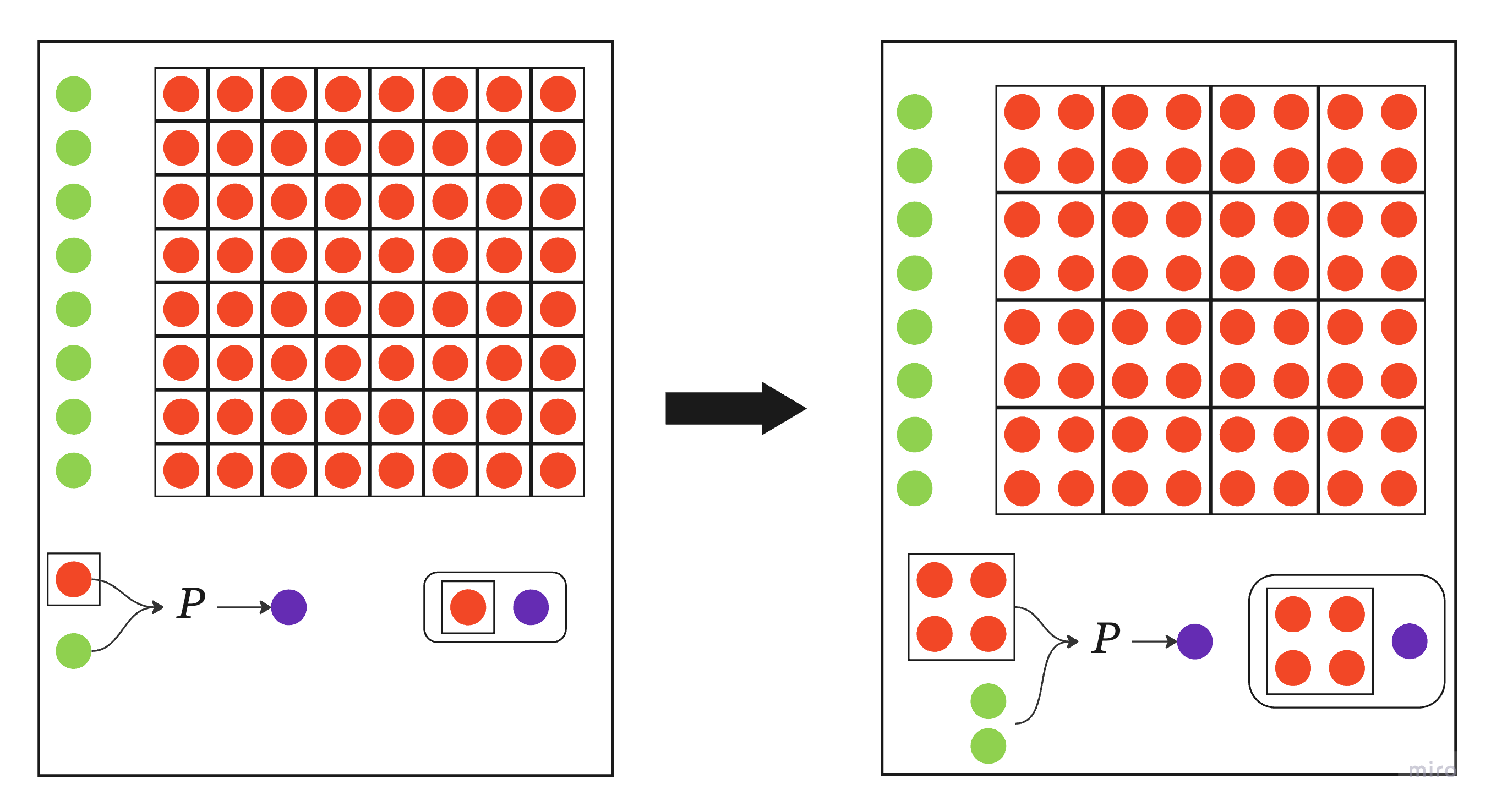}
    \caption{Baseline and multiproof layouts in the Avail instantiation.}
    \label{fig:example}
\end{figure}
\vspace{0.12em}

Figure~\ref{fig:example} illustrates the difference between the baseline per-cell layout and the grouped multiproof layout used in the Avail instantiation.

\subsection{Baseline and Multiproof Layouts}

In the baseline layout, every 32-byte data entry is stored with its own 48-byte proof.

\vspace{0.12em}
\begin{table}[htbp]
\centering
\small
\begin{tabular}{|c|c|}
\hline
\textbf{Proof} {[u8;48]} & \textbf{Data} {[u8;32]} \\
\hline
\end{tabular}
\caption{Baseline layout.}
\label{tab:baseline-layout}
\end{table}
\vspace{0.12em}

In the multiproof layout, several scalars are grouped into one authenticated object represented by an \texttt{MCell}, which stores a vector of scalars, one aggregated proof, and a \texttt{GCellBlock} describing the covered block region.

\vspace{0.12em}
\begin{table}[htbp]
\centering
\small
\begin{tabular}{|c|c|c|c|}
\hline
\textbf{Proof} & \textbf{GCellBlock} & \textbf{Count} & \textbf{Scalars} \\
{[u8;48]} & $4 \times u32$ & $u32$ & $N \times [u64;4]$ \\
\hline
\end{tabular}
\caption{Multiproof layout.}
\label{tab:multiproof-layout}
\end{table}
\vspace{0.12em}

Tables~\ref{tab:baseline-layout} and~\ref{tab:multiproof-layout} show the difference between the baseline per-entry object and the grouped authenticated object used by the multiproof path.

\subsection{Storage Analysis}

In the baseline design, each DHT object occupies
\[
32 + 48 = 80 \text{ bytes.}
\]
For 64 entries, the total is
\[
64 \times 80 = 5{,}120 \text{ bytes.}
\]

If 4 entries are grouped into one authenticated cell, each cell contains
\[
4 \times 32 = 128 \text{ bytes}
\]
plus one 48-byte proof, for a total of
\[
128 + 48 = 176 \text{ bytes per cell.}
\]
The amortized cost is therefore
\[
\frac{128 + 48}{4} = 44 \text{ bytes per entry,}
\]
and 64 entries require
\[
16 \times 176 = 2{,}816 \text{ bytes.}
\]

These calculations isolate proof amortization and do not include deployment metadata such as \texttt{GCellBlock}, scalar counts, serialization framing, DHT keys, routing metadata, replication overhead, or transport framing. They should therefore be interpreted as lower-bound proof amortization figures rather than full end-to-end on-wire storage or bandwidth costs.

\section{Implementation in Avail}
\label{sec:implementation}

We implemented the design in Avail's Rust-based light-client stack, using the open-source Avail light client repository as the baseline~\cite{availlightclient}. The multiproof integration described in this paper was merged through pull request \#778 at commit \texttt{4b17189}~\cite{availpr778}. Full nodes generate proofs and commitment context for DAS serving. Fat clients fetch these objects, convert them into grouped authenticated \texttt{MCell} objects keyed by \texttt{GCellBlock}, and publish them into the DHT. Light clients sample coordinates, fetch grouped objects, reconstruct the evaluation context, and verify multiple sampled cells against one proof. If a grouped object fails verification or cannot be retrieved within the configured timeout, the light client treats that attempt as a failed lookup and retries with another peer.

\section{Experimental Results}
\label{sec:results}

We evaluate the proposed multiproof integration along four dimensions: verifier resource consumption, network behavior, service capacity, and sensitivity to system parameters.

\subsection{Experimental Setup}

\paragraph{Hardware and software.}
Experiments were conducted on AWS c6a.\{2xlarge, 4xlarge\} instances for fat clients and m6a.large instances for light clients. All nodes used Rust 1.85, \texttt{blst} for pairing operations, and \texttt{tokio} for asynchronous execution. The DHT implementation is Kademlia over TCP with $\alpha=3$, bucket size $k=20$, concurrency window 16, and replication factor 5.

\paragraph{Environment model.}
The reported results were obtained under a benign-cloud deployment model rather than an adversarial routing model. DHT hit rate denotes successful retrieval within the experiment timeout window under this deployment. Accordingly, a reported hit rate of 100\% means that all measured lookups in that workload and time window succeeded; it is not a universal guarantee about all future deployments or adversarial conditions.

\paragraph{Datasets and workloads.}
We generated Reed--Solomon-encoded blocks from 1\,KB to 2\,MB and evaluated four configurations: \textbf{Vanilla}, which serves each sampled cell with an independent proof; \textbf{Batched Single Proofs}, which batch-verifies independent KZG openings using pairing-product checks; \textbf{Grouped Only}, which uses grouped DHT objects without multiproof aggregation; and \textbf{PMP}, which serves grouped objects using aggregated multiproofs. Unless varied explicitly, the PMP configuration used $|T_g| = 16$. Each experiment ran for 30 minutes and was repeated five times with different random seeds. Reported values are means with 95\% bootstrap confidence intervals.

\paragraph{Metrics.}
We report verifier CPU cost, peak memory usage, DHT hit rates, DHT put rates, and the deployment size needed to keep 95th percentile lookup latency below 250\,ms. CPU is reported in normalized relative units with the single-proof baseline fixed at 100 on the same hardware. Per-fat capacity denotes the maximum sustained payload served per fat client while meeting the latency target.

\subsection{Evaluation Scope and Ablation}

Table~\ref{tab:ablation} separates three sources of improvement: verifier-side batching of independent proofs, DHT-side grouping of retrieval objects, and full multiproof aggregation. Batched verification closes part of the verifier-cost gap relative to the naive per-proof baseline, but it does not reduce proof bytes or DHT object count. Grouped retrieval improves locality and storage fragmentation. The best end-to-end gains appear when proof aggregation and grouped transport are combined.

\vspace{0.12em}
\begin{table}[htbp]
\centering
\footnotesize
\setlength{\tabcolsep}{4pt}
\resizebox{\linewidth}{!}{%
\begin{tabular}{@{}lcccc@{}}
\toprule
\textbf{Metric} & \textbf{Vanilla} & \textbf{Batched} & \textbf{Grouped} & \textbf{Grouped + PMP} \\
\midrule
Verifier CPU @1{,}000 batch & 110 & 82 & 92 & 60 \\
Peak memory (GB) & 1.2 & 1.05 & 1.0 & 0.7 \\
DHT hit rate @1\,KB & 78\% & 78\% & 88\% & 100\% \\
DHT hit rate @1.5--2\,MB & 12\% & 12\% & 48\% & 98\% \\
Fats to meet SLO (1\,KB) & 40 & 36 & 18 & 10 \\
Per-fat capacity (B/fat @2\,MB) & 16 & 19 & 46 & 80 \\
\bottomrule
\end{tabular}%
}
\caption{Ablation across independent-proof batching, grouped retrieval, and polynomial multiproofs.}
\label{tab:ablation}
\end{table}
\vspace{0.12em}

\subsection{Verifier Cost}

\begin{figure}[H]
\centering
\includegraphics[width=0.56\linewidth]{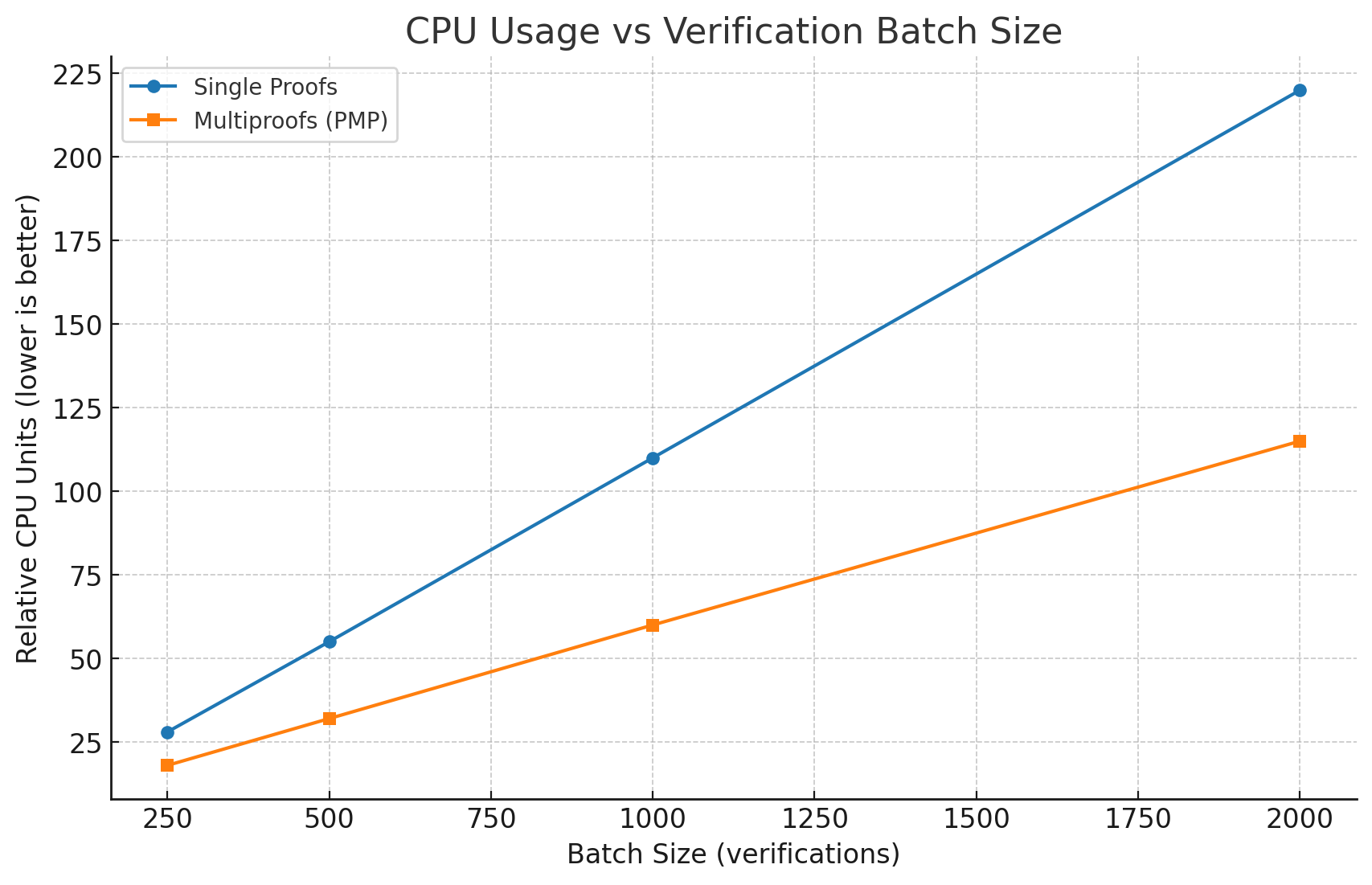}
\caption{CPU usage by batch size.}
\label{fig:cpu-line}
\end{figure}

\begin{figure}[H]
\centering
\includegraphics[width=0.44\linewidth]{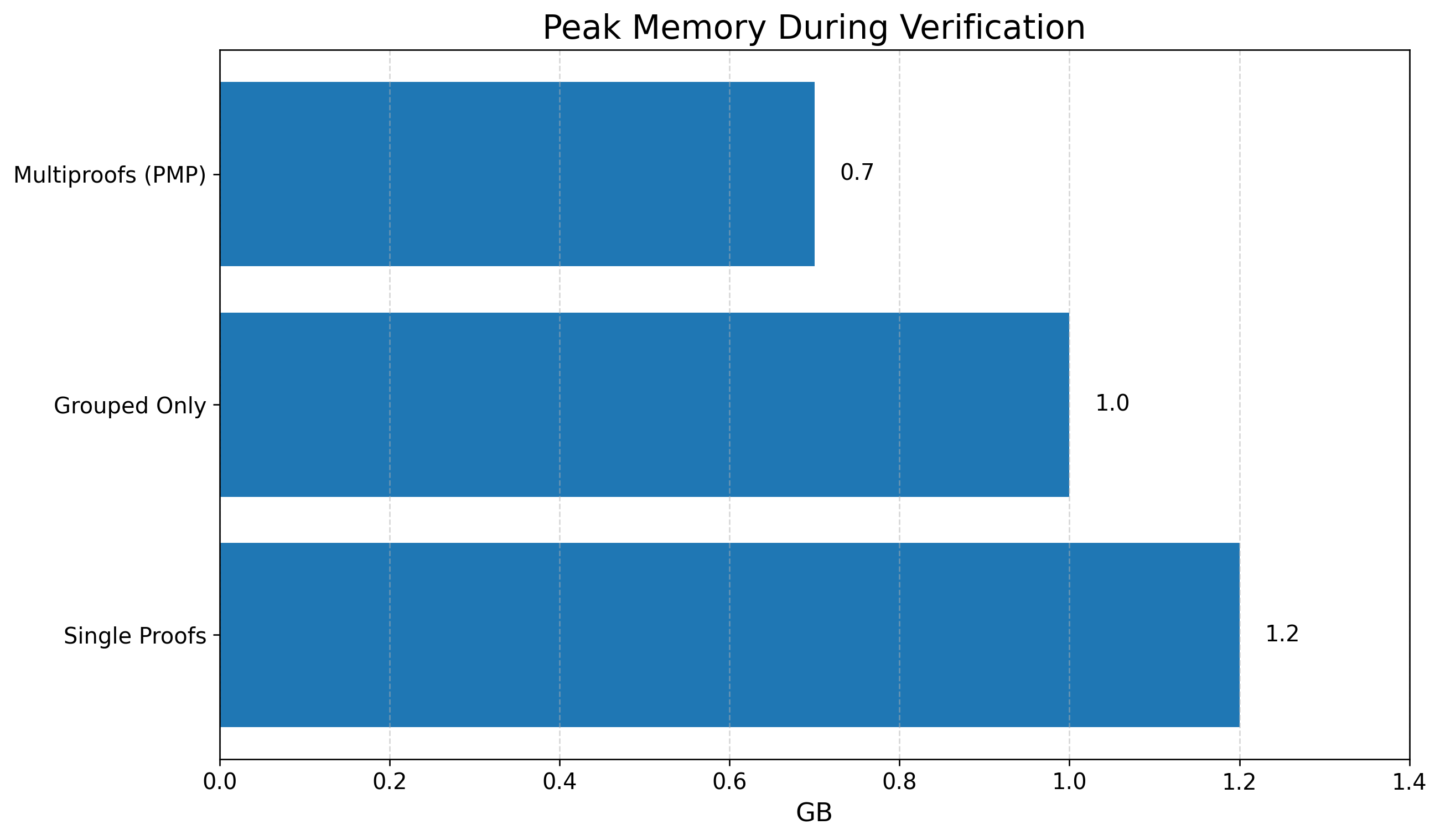}
\caption{Peak memory usage across the baseline and ablation configurations.}
\label{fig:mem-hbar}
\end{figure}

Figure~\ref{fig:cpu-line} shows that PMP lowers verifier CPU cost across batch sizes. At a batch size of 1{,}000 verifications, PMP uses about 60 normalized CPU units compared with about 110 for the single-proof baseline. Batched independent proofs reduce this to about 82 units, which narrows but does not eliminate the advantage of PMP. The same trend appears in memory usage in Figure~\ref{fig:mem-hbar}.

\vspace{0.12em}
\vspace{0.12em}

\vspace{0.12em}
\vspace{0.12em}

\subsection{Network Behavior}

\begin{figure}[H]
\centering
\includegraphics[width=0.56\linewidth]{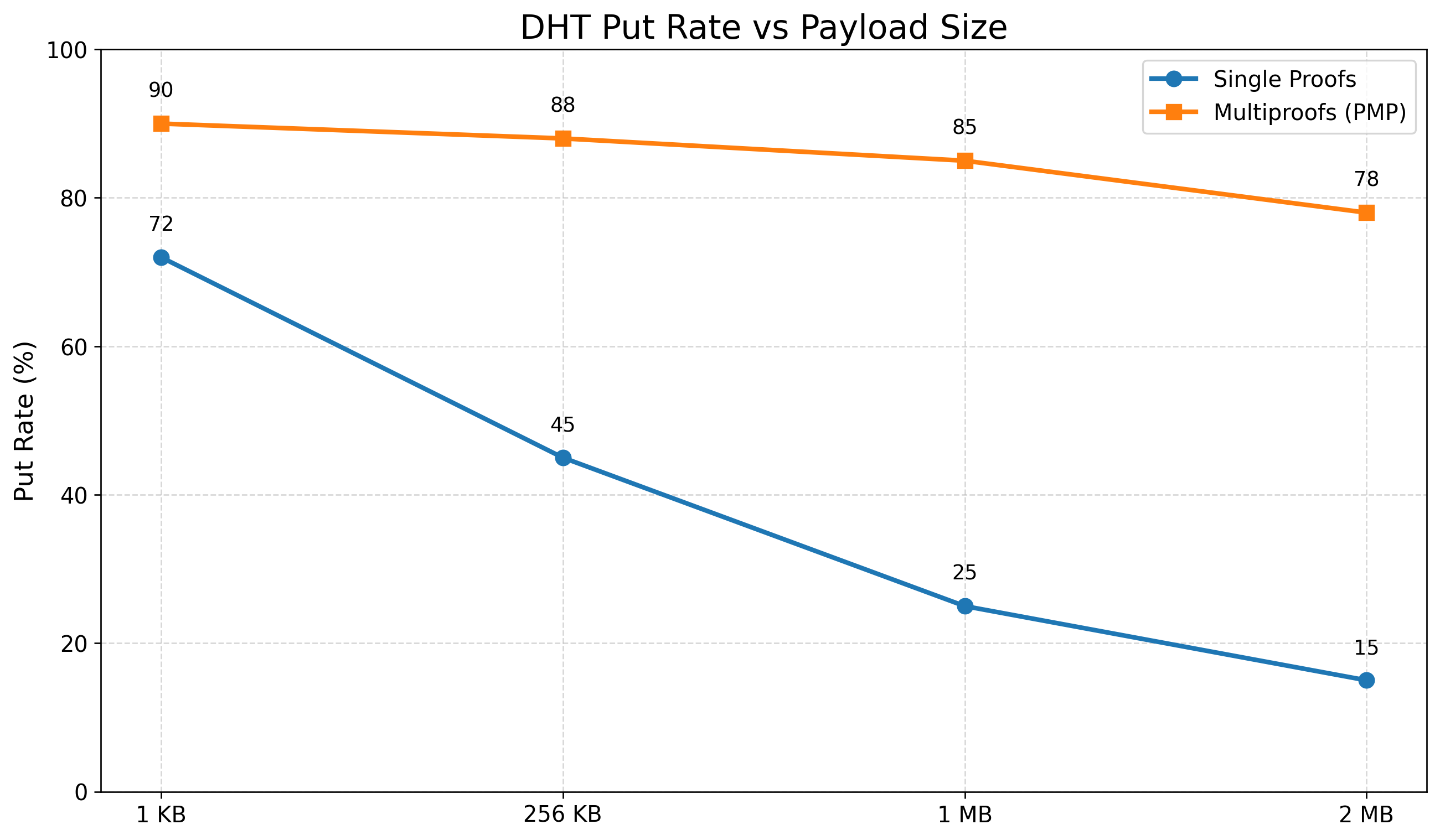}
\caption{DHT put rates by payload size for the baseline and PMP configurations.}
\label{fig:put-rates-line}
\end{figure}

\begin{figure}[H]
\centering
\includegraphics[width=0.56\linewidth]{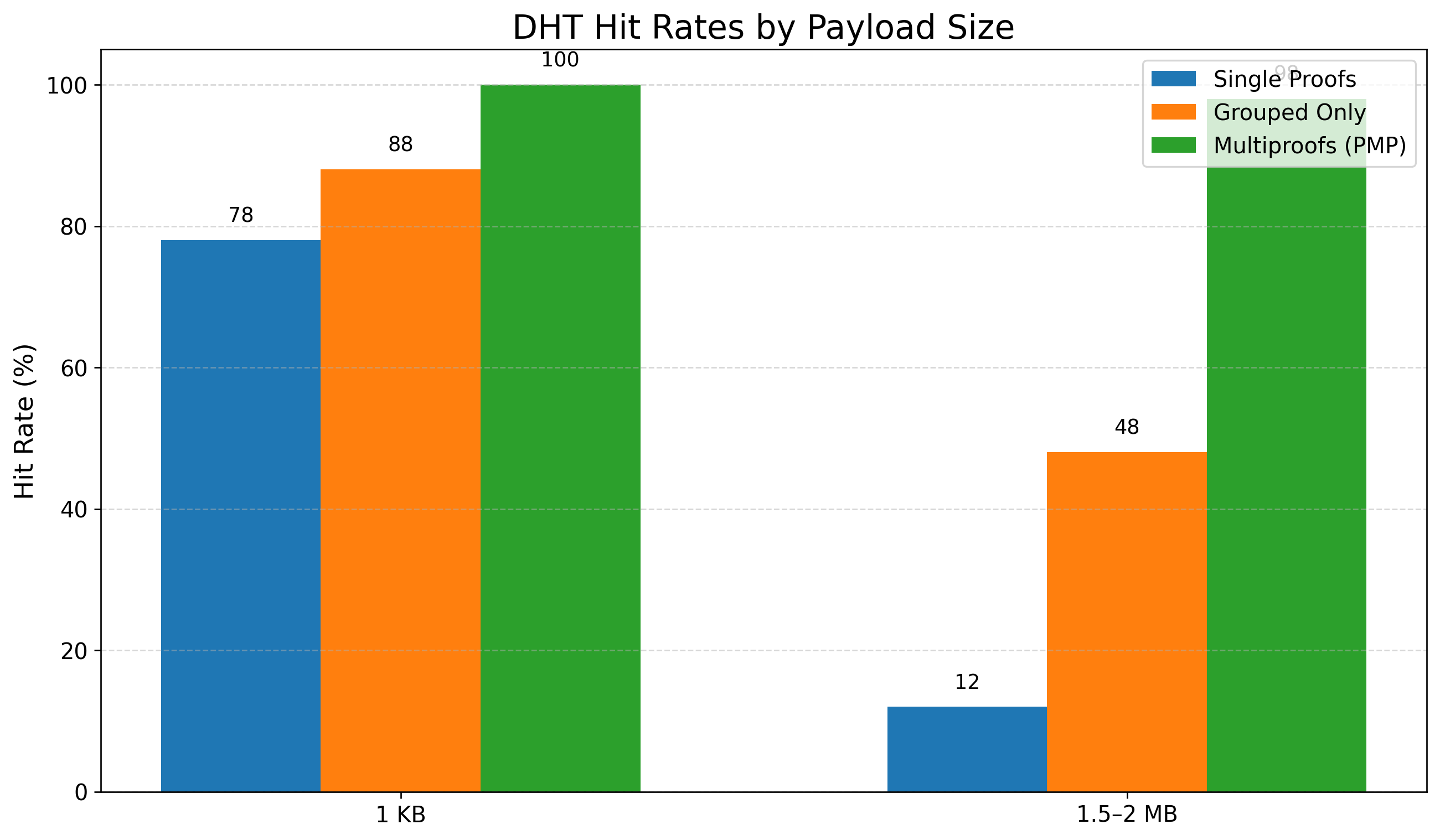}
\caption{DHT hit rates by payload size for Vanilla, Grouped Only, and Multiproofs.}
\label{fig:hit-rates-grouped}
\end{figure}

Figure~\ref{fig:put-rates-line} shows that PMP sustains higher DHT put rates as payload size increases. Figure~\ref{fig:hit-rates-grouped} shows that DHT hit rates improve much more substantially once grouped retrieval and multiproofs are introduced. At 1\,KB payload size, hit rate rises from 78\% in the vanilla design to 88\% with grouped retrieval alone and 100\% with PMP. At 1.5--2\,MB payload sizes, the hit rate rises from 12\% to 48\% with grouping and to 98\% with PMP. Ordinary batched verification does not materially improve DHT hit or put behavior, because it changes only verifier arithmetic and not the number or layout of transported objects.

\vspace{0.12em}
\vspace{0.12em}

\vspace{0.12em}
\vspace{0.12em}

\subsection{Capacity and Sensitivity}

\begin{figure}[H]
\centering
\includegraphics[width=0.58\linewidth]{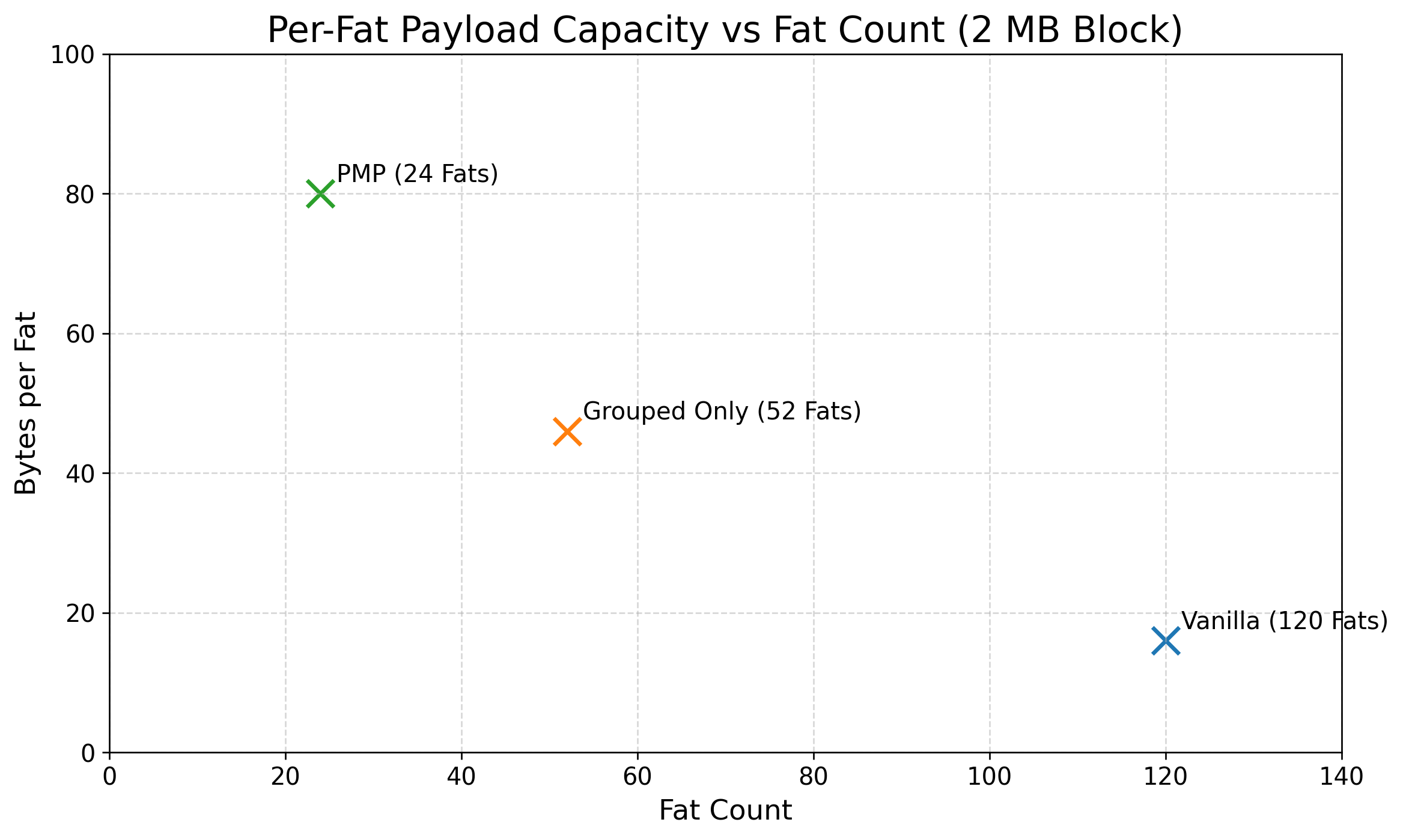}
\caption{Per-fat payload capacity versus fat count at 2\,MB block size.}
\label{fig:capacity-scatter}
\end{figure}

\begin{figure}[H]
\centering
\includegraphics[width=0.50\linewidth]{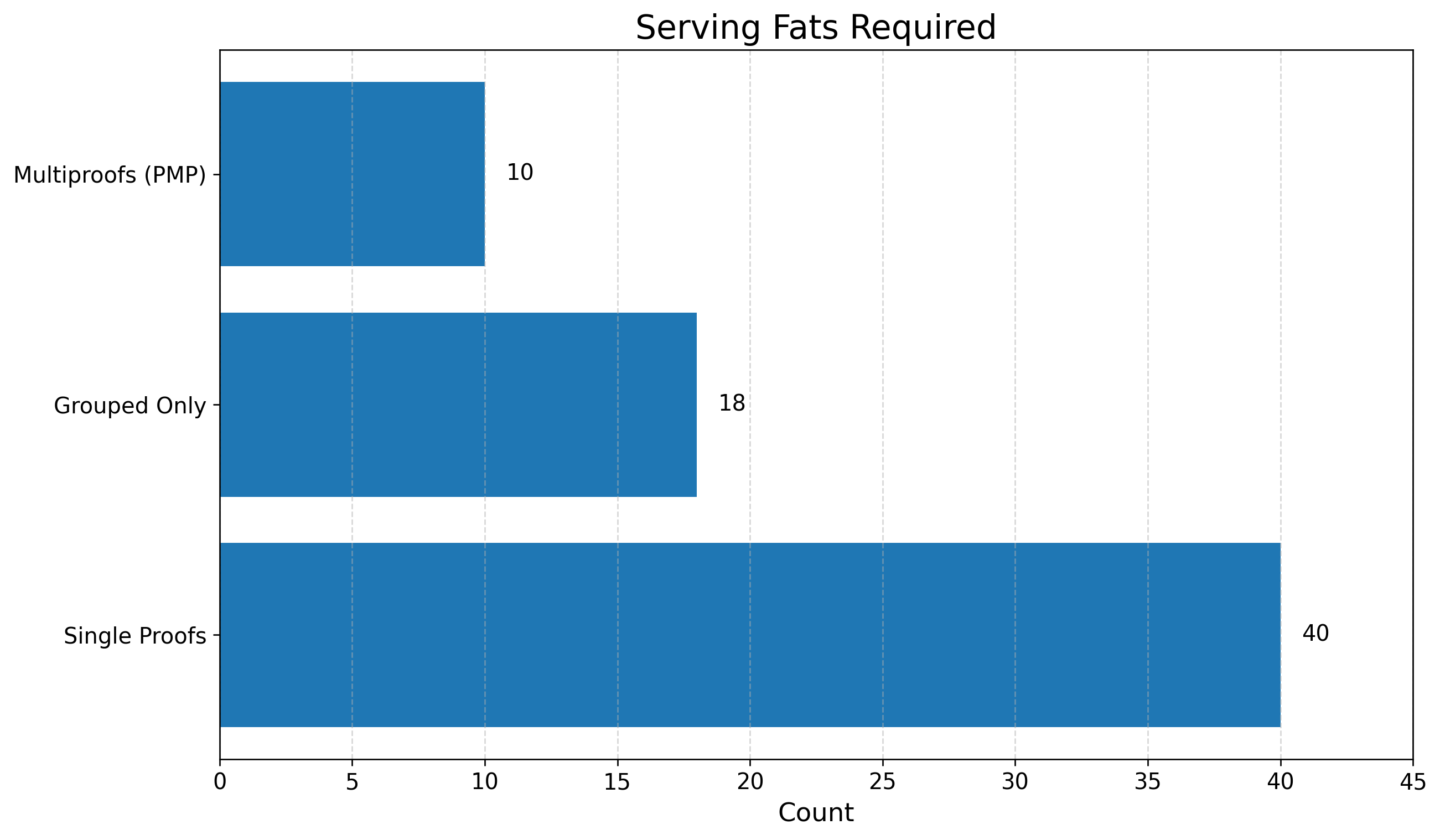}
\caption{Serving fats required across configurations.}
\label{fig:fats-hbar}
\end{figure}

\begin{figure}[H]
\centering
\includegraphics[width=0.50\linewidth]{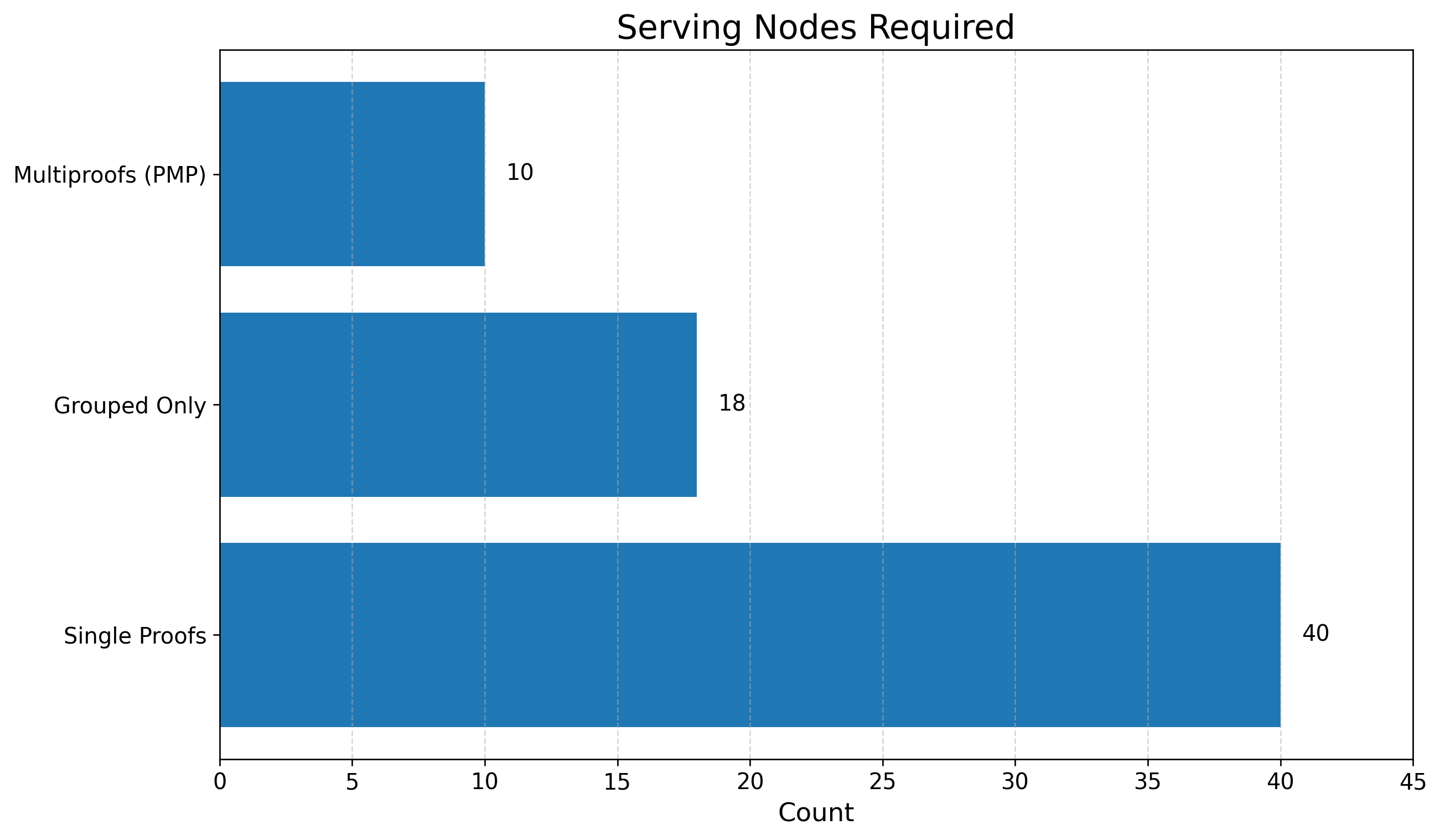}
\caption{Servers required across the baseline and PMP configurations.}
\label{fig:servers-hbar}
\end{figure}

Figure~\ref{fig:capacity-scatter} shows the improvement in per-fat payload capacity at 2\,MB block size. Figure~\ref{fig:fats-hbar} shows the reduction in serving fat clients across configurations, and Figure~\ref{fig:servers-hbar} shows the corresponding reduction in server requirements. For the 1\,KB payload scenario, the number of serving fat clients required to satisfy the latency objective drops from 40 in the vanilla system to 36 with batched single proofs, 18 with grouped retrieval alone, and 10 with PMP. At larger payloads, per-fat capacity improves from 16 to 19\,B/fat under batched independent verification, to 46\,B/fat under grouped retrieval, and to 80\,B/fat with PMP.

\vspace{0.12em}
\vspace{0.12em}

\vspace{0.12em}
\vspace{0.12em}

\vspace{0.12em}
\vspace{0.12em}

Varying $|T_g| \in \{8,16,32\}$ shows a convex trade-off between proving efficiency and verifier interpolation cost, with $|T_g| = 16$ giving the best end-to-end latency. Under 10\%--30\% DHT churn, PMP maintains hit rates above 75\% at 2\,MB payload size, while the baseline falls below 20\%.

\subsection{Infrastructure Cost Model}

We define infrastructure cost as the cloud compute cost of the minimum deployment required to satisfy the target lookup latency objective. In the experiments, this corresponds to the number and type of fat client instances needed to keep the 95th percentile lookup latency below the target threshold.

\vspace{0.12em}
\begin{table}[htbp]
\centering
\small
\begin{tabular}{@{}lccc@{}}
\toprule
\textbf{Metric} & \textbf{Vanilla} & \textbf{PMP} & \textbf{Relative} \\
\midrule
Verifier CPU @1{,}000 batch & 110 & 60 & $\downarrow45\%$ \\
Peak memory (GB) & 1.2 & 0.7 & $\downarrow42\%$ \\
DHT hit rate @1\,KB & 78\% & 100\% & $\uparrow1.28\times$ \\
DHT hit rate @1.5--2\,MB & 12\% & 98\% & $\uparrow8.17\times$ \\
Fats to meet SLO (1\,KB) & 40 & 10 & $\downarrow75\%$ \\
Per-fat capacity (B/fat @2\,MB) & 16 & 80 & $\uparrow5\times$ \\
Infra compute cost & --- & --- & $\downarrow\le45\%$ \\
\bottomrule
\end{tabular}
\caption{Results summary comparing the baseline design with the PMP deployment.}
\label{tab:summary}
\end{table}
\vspace{0.12em}

Table~\ref{tab:summary} consolidates the main end-to-end gains of PMP relative to the baseline design. It should be read together with Table~\ref{tab:ablation}, which shows that part of the gain comes from grouped transport and part from proof aggregation itself.

\subsection{Reproducibility}

All experiments fix compiler versions, random seeds, and network parameters. Metrics are exported through Prometheus, and confidence intervals are computed using bootstrap resampling. CPU usage is reported as normalized utilization relative to the baseline single-proof configuration on the same hardware. Because CPU and memory are reported in normalized units on fixed hardware, the results support relative comparison across configurations but do not by themselves establish feasibility on mobile- or browser-class devices.

\FloatBarrier
\section{Conclusion}
\label{sec:conclusion}

This paper studied polynomial multiproofs as a systems optimization for blockchain light clients and validated the design through an implementation and evaluation in Avail. By aggregating multiple sampled evaluations into a single proof and embedding that representation directly into the DHT workflow, the proposed design reduces proof redundancy, improves locality, and lowers verifier overhead.

The results show lower CPU and memory consumption, higher hit rates under large payloads, and lower infrastructure requirements, while also making explicit the trade-off introduced by grouped retrieval under adversarial or privacy-sensitive conditions. The design is complementary to stronger dissemination substrates such as PANDAS: our contribution targets proof amortization and retrieval locality within the authenticated data path.

\FloatBarrier
\vspace{-0.15em}
\section*{Nomenclature}

\begin{itemize}
  \item \textbf{DAS}: Data Availability Sampling
  \item \textbf{DHT}: Distributed Hash Table
  \item \textbf{KZG}: Kate--Zaverucha--Goldberg polynomial commitment
  \item \textbf{PCS}: Polynomial Commitment Scheme
  \item \textbf{PMP}: Polynomial Multiproof
  \item \textbf{SRS}: Structured Reference String
  \item $\mathbb{F}$: finite field
  \item $\mathbb{G}_1, \mathbb{G}_2$: pairing groups
  \item $e(\cdot,\cdot)$: bilinear pairing
  \item $T_g$: shared evaluation micro-domain
  \item $Z_{T_g}$: vanishing polynomial over $T_g$
\end{itemize}

\vspace{0.35em}

{\footnotesize
\raggedright
\setlength{\itemsep}{0pt}
\setlength{\parskip}{0pt}

}

\begin{thebibliography}{99}

\bibitem{albassam2018fraud}
Mustafa Al-Bassam, Alberto Sonnino, and Vitalik Buterin.
\newblock Fraud and Data Availability Proofs: Maximising Light Client Security and Scaling Blockchains with Dishonest Majorities.
\newblock \textit{arXiv preprint arXiv:1809.09044}, 2018.

\bibitem{albassam2019lazyledger}
Mustafa Al-Bassam.
\newblock LazyLedger: A Distributed Data Availability Ledger With Client-Side Smart Contracts.
\newblock \textit{arXiv preprint arXiv:1905.09274}, 2019.

\bibitem{hallandersen2023foundations}
Mathias Hall-Andersen, Mark Simkin, and Benedikt Wagner.
\newblock Foundations of Data Availability Sampling.
\newblock \textit{IACR Cryptology ePrint Archive}, Paper 2023/1079, 2023.

\bibitem{krol2023ethereumdas}
Micha{\l} Kr{\'o}l, Onur Ascigil, Sergi Rene, Etienne Rivi{\`e}re, Matthieu Pigaglio, Kaleem Peeroo, Vladimir Stankovic, Ramin Sadre, and Felix Lange.
\newblock Data Availability Sampling in Ethereum: Analysis of P2P Networking Requirements.
\newblock \textit{arXiv preprint arXiv:2306.11456}, 2023.

\bibitem{kate2010kzg}
Aniket Kate, Gregory M. Zaverucha, and Ian Goldberg.
\newblock Constant-Size Commitments to Polynomials and Their Applications.
\newblock In \textit{Advances in Cryptology -- ASIACRYPT 2010}, pages 177--194. Springer, 2010.

\bibitem{boneh2020multipoint}
Dan Boneh, Justin Drake, Ben Fisch, and Ariel Gabizon.
\newblock Efficient Polynomial Commitment Schemes for Multiple Points and Polynomials.
\newblock \textit{IACR Cryptology ePrint Archive}, Report 2020/081, 2020.

\bibitem{maymounkov2002kademlia}
Petar Maymounkov and David Mazieres.
\newblock Kademlia: A Peer-to-Peer Information System Based on the XOR Metric.
\newblock In \textit{Proceedings of IPTPS}, pages 53--65. Springer, 2002.

\bibitem{pigaglio2025pandas}
Matthieu Pigaglio, Onur Ascigil, Micha{\l} Kr{\'o}l, Sergi Rene, Felix Lange, Kaleem Peeroo, Ramin Sadre, Vladimir Stankovic, and Etienne Rivi{\`e}re.
\newblock PANDAS: Peer-to-peer, Adaptive Networking for Data Availability Sampling within Ethereum Consensus Timebounds.
\newblock \textit{arXiv preprint arXiv:2507.00824}, 2025.

\bibitem{balduf2023ipfs}
Lennart Balduf, Daniel Trautwein, Christopher Stransky, Florian Tschorsch, and Tobias Mueller.
\newblock Investigating the Decentralization of IPFS.
\newblock In \textit{Proceedings of the 2023 ACM on Internet Measurement Conference}, pages 599--612, 2023.

\bibitem{availlightclient}
Avail Project.
\newblock Avail Light Client.
\newblock GitHub repository.
\newblock \url{https://github.com/availproject/avail-light}.

\bibitem{availpr778}
Avail Project.
\newblock Multiproof integration for Avail light client.
\newblock GitHub pull request \#778, commit \texttt{4b1718960bfc956a937de34fe3f609c7dbfee439}.
\newblock \url{https://github.com/availproject/avail-light/pull/778}.

\end{thebibliography}
\end{document}